\documentstyle[12pt,preprint,aps]{revtex}

\def \to {\rightarrow}
\def \beq {\begin{equation}}
\def \eeq {\end{equation}}
\def \ba {\begin{eqnarray}}
\def \ea {\end{eqnarray}}
\def \jpsi {J/\psi}
\def \mtss {\langle 0 |{\cal O } ^{J/\psi}_{1} [ ^3S_1 ]| 0 \rangle}
\def \mtso {\langle 0 |{\cal O } ^{J/\psi}_{8} [ ^3S_1 ]| 0 \rangle}
\def \moso {\langle 0 |{\cal O } ^{\jpsi}_{8} [ ^1S_0 ]| 0 \rangle}
\def \mtpo {\langle 0 |{\cal O } ^{\jpsi}_{8} [ ^3P_0 ]| 0 \rangle}
\def \mtpj {\langle 0 |{\cal O } ^{\jpsi}_{8} [ ^3P_J ]| 0 \rangle}

\begin{document}
\draft
\title{Associated $ \jpsi + \gamma $ production through double Pomeron
exchange :  The nature
of the Pomeron and  hard diffractive factorization breaking}
\author{Jia-Sheng Xu}
\address{Department of Physics, Peking University, Beijing 100871, China}
\author{Hong-An Peng}
\address{ China Center of Advance Science and
Technology (World Laboratory), Beijing 100080, China \\
and Department of Physics, Peking University, Beijing 100871,
 China }
\maketitle

\begin{abstract}
We present a study of associated $\jpsi + \gamma$ production
through double Pomeron exchange at energies reached at the Fermilab
Tevatron and CERN LHC based on the
Ingelman-Schlein model for hard diffractive scattering and the factorization
formalism of nonrelativistic QCD for quarkonia production. We find that this process
$(p + {\bar p} \to p + {\bar p} + \jpsi + \gamma + X)$ can be used to
probe the gluon
content of the Pomeron and study the nature of hard diffractive 
factorization breaking.
\vskip 3mm
\end{abstract}

\vfill\eject\pagestyle{plain}\setcounter{page}{1}

%\twocolumn
\narrowtext

\par
In high-energy strong interactions the Regge trajectory with a vacuum quantum
number, the Pomeron, plays a particular and very important role in soft
processes in hadron-hadron collisions \cite{collins}.
However, the nature of the Pomeron and its interaction with hadrons
remain a mystery.
Ingelman and Schlein \cite{ingelmam} pointed out that  hard diffractive
scattering processes would give new and valuable insight about the nature
of the Pomeron. It is assumed that the Pomeron, similar to the nucleon,
is composed of partons, mainly of gluons, and the hard diffraction can be
studied in a factorized way:
First, the Pomeron is emitted from the diffractively scattered hadron, then
one parton of the Pomeron takes part in hard subprocesses.
Therefore, the partonic structure of the Pomeron  could be studied
experimentally.
The Pomeron has a partonic structure was confirmed
subsequently by  various experiments \cite{ua8,zeus,h1,cdfwj}.

\par
However, there is an important problem, the factorization problem.
The Ingelman and Schlein (IS) model for hard diffraction is based on the
assumption of hard diffractive scattering factorization.
Recently, a factorization theorem
has been proven by Collins \cite{jcollins} for the lepton induced
hard diffractive scattering processes,
such as diffractive deep inelastic scattering (DDIS) and diffractive
direct photoproduction of jets.
In contrast,
no factorization theorem has been established for hard diffraction
in hadron-hadron collisions.
At large $|t|$ ($t$ is the square of the hadron's four-momentum transfer),
where perturbative QCD applies to the Pomeron, it has been proven that
there is a leading twist contribution which breaks the factorization
theorem for hard diffraction in hadron-hadron collision \cite{copom}.
This coherent
hard diffraction was observed by the UA8 Collaboration in diffractive jet
production, in this experiment  $t$ is in the region $ -2 \leq t \leq 
-1 $ GeV$^2$ \cite{ua8}. On the other hand, for the Pomeron at small $|t|$
, nonperturbative QCD dominates, it is unclear whether hard diffractive
factorization is valid or not in hadron-hadron collisions.
In phenomenology,
the large discrepancy between the theoretical prediction and
the Tevatron date on 
the diffractive production of
jets and weak bosons, {\it at al.}, signals
a breakdown of hard diffractive factorization in hadron-hadron
collisions \cite{testa}, 
but the nature of hard diffrative factorization breaking is still unclear.

\par
In a previous paper, we have studied associated $\jpsi + \gamma $ production
in single diffractive (SD) scattering at the Fermilab Tevatron and CERN LHC
energies based
on the IS model for hard diffractive scattering and
the nonrelativistic QCD (NRQCD) factorization scheme for quarkonia
production \cite{xu}.
In this BRIEF REPORT,
we complete our study of associated $\jpsi + \gamma $ diffractive production
by discuss associated $\jpsi + \gamma $ production at large $P_T$ through
double Pomeron exchange (DPE):
\beq
\label{proc}
p(P_p) + {\bar p}(P_{\bar p}) \to p(P^{\prime}_{p}) +
{\bar p}(P^{\prime}_{\bar p}) + \jpsi (P) + \gamma (k) + X.
\eeq
This process is of special interesting  because the large $P_T $
$\jpsi $ produced is in the central rapidity region and  is easy to be
detected through its leptonic decay modes,  and the
$\jpsi $'s large $P_T $ is balance by the associated high energy photon.
We will see the measurement of this process
at the Tevatron and LHC would shed light on the nature of the Pomeron
and the hard diffractive factorization breaking. Furthermore, this process
is also interesting to the study of heavy quarkonium production mechanism,
and the $P_T$ smearing effects.

\par
Within the NRQCD framework \cite{nrqcd,power} $\jpsi $ is described in terms
of Fock state decompositions as
\ba
|\jpsi \rangle &=& O(1)~ |c{\bar c}[^3S_{1}^{(1)}] \rangle +
            O(v) |c{\bar c}[^3P_{J}^{(8)}] g \rangle  \nonumber \\
        & & + O(v^2) |c{\bar c}[^1S_{0}^{(8)}] g \rangle +
            O(v^2) |c{\bar c}[^3S_{1}^{(1,8)}] g g \rangle  \nonumber \\
        & & + O(v^2) |c{\bar c}[^3P_{J}^{(1,8)}] g g\rangle + \cdots ,
\ea
where the $c{\bar c}$ pairs are indicated within the square brackets in
spectroscopic notation. The pairs' color states are indicate by singlet (1)
or octet (8) superscripts. The color octet $c{\bar c}$ states can make
a transition into a physical $\jpsi$ state by soft chromoelectric
dipole ($E1$) transition(s) or chromomagnitic dipole ($M1$) transition(s)
\beq
(c {\bar c})[^{2S + 1}L_{j}^{(8)}] \to \jpsi  .
\eeq
These color-octet contributions are essential for cancelling the logarithmic
infrared divergences which appear in the color-singlet model calculations of
the production cross sections and annihilation decay rates for P-wave
charmonia,
and for solving the $\psi^{\prime}$
and direct $\jpsi$ ``surplus'' problems at the Fermilab
Tevatron\cite{cdfjpsi,brt-cho}.
The NRQCD factorization scheme \cite{nrqcdfs} has
been established to systematically separate high- and low- energy scale
interactions.
Furthermore,  NRQCD power counting rules can be exploited to determine the
dominant contributions to various quarkonium processes\cite{power}.
For direct $\jpsi $ production, the color-octet matrix elements,
$\mtso$ , $\moso$ and $ \mtpj $ are all scaling as $m_c^3 v_c^7 $. So these
color-octet contributions to $\jpsi $ production must be included for
consistency.

\par
On the partonic level, associated $\jpsi + \gamma $
production are composed of the gluon fusion:
\ba
g_1 (p_1) + g (p_2) & \to & \gamma + (c{\bar c})[^3S_{1}^{(1)},
^3S_{1}^{(8)}]~,  \nonumber \\
g_1 (p_1) + g (p_2) & \to & \gamma + (c{\bar c})[^1S_{0}^{(8)},
^3P_{J}^{(8)}]~ .
\ea
The quark initiated subprocesses ($q{\bar q}$ channel) are strongly
suppressed and will be neglected further. The color-singlet gluon-gluon
fusion contribution to associated $\jpsi + \gamma $ production is well
known \cite{berg}:
\ba
\label{sg3s11}
\frac{d\hat{\sigma}} {d\hat{t}}({\rm singlet}) &=&
\frac{1}{16 \pi \hat{s}^2} \overline{\Sigma }
    |M(g + g \to \gamma + (c{\bar c})[^3S_{1}^{(1)}] )|^2  \nonumber \\
 &~&   \times \frac{1}{18 m_c} \mtss ,
\ea

\noindent
where $\hat{s} = (p_1 + p_2)^2$, $\mtss $ is
the color-singlet matrix element which is related to the lepton
decay width of $\jpsi$ in NRQCD 
\beq
\mtss = \frac{\Gamma (\jpsi \to j^+ l^-)  m_c^2 }{(2/9) \pi e_c^2 \alpha^2 },
\eeq
\noindent
where 
$e_c = 2/3$.

\par
The average-squared amplitude of the subprocess
$ g + g  \to  \gamma + (c{\bar c})[^3S_{1}^{(8)}]$ can be obtained from the
average-squared amplitude of
$g + g  \to  \gamma + (c{\bar c})[^3S_{1}^{(1)}]$ by taking into account the
defferent  color factor. The result is
\ba
\label{sg3s18}
~&&\frac{d\hat{\sigma}} {d\hat{t}}[g + g  \to
   \gamma + (c{\bar c})[^3S_{1}^{(8)}]  \to \gamma + \jpsi]  \nonumber \\
~&&~~~~= \frac{1}{16 \pi \hat{s}^2 } \frac{15}{6}
\overline{\Sigma }
    |M(g + g \to \gamma + (c{\bar c})[^3S_{1}^{(1)}] )|^2 \nonumber \\
~&&~~~~~    \times \frac{1}{24 m_c} \mtso .
\ea
The average-squared amplitudes of the subprocesses
$g + g  \to  \gamma + (c{\bar c})[^1S_{0}^{(8)}]$ and
$g + g  \to  \gamma + (c{\bar c})[^3P_{J}^{(8)}]$ can be found in \cite{kim},

\ba
\label{sgsp08}
~&&\frac{d\hat{\sigma}} {d\hat{t}}[g + g  \to
   \gamma + (c{\bar c})[^1S_{0}^{(8)}]  \to \gamma + \jpsi] \nonumber \\
~&&~~~~= \frac{1}{16 \pi \hat{s}^2 } \overline{\Sigma }
    |M(g + g \to \gamma + (c{\bar c})[^1S_{0}^{(8)}] )|^2  \nonumber \\
~&&~~~~~    \times \frac{1}{8 m_c} \moso , \nonumber \\ 
~&&\frac{d\hat{\sigma}} {d\hat{t}}[g + g  \to
   \gamma + (c{\bar c})[^3P_{J}^{(8)}]  \to \gamma + \jpsi] \nonumber \\
~&&~~~~= \frac{1}{16 \pi \hat{s}^2 } \sum_{J} \overline{\Sigma }
    |M(g + g \to \gamma + (c{\bar c})[^3P_{J}^{(8)}] )|^2  \nonumber \\
~&&~~~~~    \times \frac{1}{8 m_c} \mtpo ,
\ea
where the heavy quark spin symmetry
\beq
\mtpj = (2 J + 1 ) \mtpo
\eeq
is exploited.

\par
Now we consider the $P_T $ distribution of $\jpsi $ produced in
process (\ref{proc}).
Assuming hard diffractive factorization, the differential cross section can
be expressed as
\ba
\label{dsigma}
d\sigma &=& f_{{\rm I\!P}/p }(\xi_1) f_{{\rm I\!P}/{\bar p}}(\xi_2)
          f_{g/{\rm I\!P} }(x_1, Q^2)
          f_{g/{\rm I\!P} }(x_2, Q^2)    \nonumber \\
 &~&     \frac{d\hat{\sigma}}{d\hat{t}}[{\rm singlet + octet}]
           d\xi_1 d\xi_2 dx_1 dx_2 d\hat{t} ,
\ea
where $\xi_1 (\xi_2) $ is the momentum fraction of the proton (antiproton)
carried by the Pomeron, $f_{g/{\rm I\!P} }(x_1, Q^2)$ is the gluon
distribution function of the Pomeron, and $f_{{\rm I\!P}/p }(\xi)$
is the Pomeron flux factor integrated over $t$, 
\ba
f_{{\rm I\!P}/p }(\xi) &=& \int _{-1}^{0} f_{{\rm I\!P}/p }(\xi,t) dt,
 \nonumber \\
f_{{\rm I\!P}/p }(\xi,t) &=& \frac{\beta_1^2 (0) }{16 \pi} \xi ^{
      1 - 2 \alpha (t) } F^2(t)    \nonumber \\
~&=& K \xi ^{1 - 2 \alpha (t) } F^2(t),
\ea
where the parameters are chosen as \cite{goulianos}   \\
$$
K = 0.73~ {\rm GeV}^2,
~~\alpha (t) = 1 + 0.115 + 0.26~ ({\rm GeV}^{-2})~ {\large t} ,  $$
\beq
~F^2(t) = e^{4.6 t} ~~({\rm valid ~at } ~| t | \leq 1~ {\rm GeV}^2 ).
\eeq

\par
For numberical predictions,
we use $m_c = 1.5~{\rm GeV}, \Lambda_4 = 235~ {\rm MeV} $, and set the
factoriztion scale and the renormalization scale both equal to the
transverse mass of $\jpsi$, {\it i.e.},
$Q^2 = m_T^2 = (m_{\psi}^2 + P_T^2)$, where $P_T$ is the transverse momentum
of $\jpsi$. For the color-octet matrix elements
$\mtso , \moso$, and $\mtpo $ we use the values determined by
Beneke and Kr$\ddot{a}$mer \cite{beneke} from fitting the direct
$\jpsi$ production data at the Tevatron \cite{cdfjpsi} using GRV LO parton
distribution functions\cite{grv}: \\
$$ \mtso = 1.12 \times 10^{-2} ~{\rm GeV}^3 ,$$  
\beq
\label{matrix}
\moso + \frac{3.5}{m_c^2} \mtpo = 3.90 \times 10^{-2} ~{\rm GeV}^3 .
\eeq
The matrix elements $\moso $ and $\mtpo$ are not determined separately,
so we present the two extreme values allowed by Eq.(\ref{matrix}) as
\ba
^1S_0 ~{\rm saturated ~case:} &\moso &= 3.90 \times 10^{-2} ~{\rm GeV}^3
   \nonumber \\
~                        &\mtpo &= 0 ,   \nonumber \\
^3P_J ~{\rm saturated ~case:} &\mtpo &= 1.11 \times 10^{-2} m_c^2
~{\rm GeV}^3 \nonumber \\
~                        &\moso &= 0 .
\ea
We use the hard gluon distribution function for the Pomeron \cite{cdfwj}:
\beq
x f_{g/{\rm I\!P}} (x, Q^2) = f_g 6 x (1-x) , ~f_g = 0.7 . 
\eeq
We neglect any $Q^2$ evolution of the gluon density of the Pomeron at present
stage.  As usual, in order to suppress the Reggon contributions, we set
$\xi_1, \xi_2 \leq 0.05$.

\par
With all ingredients set as above, the $P_T$ distribution of $\jpsi $
can be calculated in a standard way from Eq. (\ref{dsigma}).
In Fig. 1 we show the $P_T$ distribution  $ B d\sigma /d P_T $ for
associated $\jpsi + \gamma$
production through DPE in $p{\bar p}$ collisions at $\sqrt{s} = 1.8 $TeV ,
integrated over the pseudo-rapidity region $-1 \leq \eta \leq 1$
(the central region). Where $ B = 0.0594$ is the $\jpsi \to \mu^+ \mu^- $
leptonic decay branching ratio.
The lowest solid line is the color-singlet gluon-gluon fusion contribution,
while the lowest dashed, dotted lines are
$ ^1 S_{0}^{(8)}$-saturated, and $^3S_{1}^{(8)}$ color-octet contributions
respectively.
The $^3P_{J}^{(8)}$-saturated color-octet contribution is almost the same
as the $ ^1 S_{0}^{(8)}$-saturated case, so we don't show it in the figure,
and we don't consider it further.
For comparision, we also show the $P_T$ distribution of $\jpsi $
produced in the inclusive process
$ p + {\bar p} \to \jpsi + \gamma + X $ and the single diffractive
process $ p + {\bar p} \to {\bar p } + \jpsi + \gamma + X $ in the same
kinematic region, the results are shown
as the upper and middle lines of Fig. 1. The code for the lines is
the same as the DPE case. As shown in Fig. 1, 
the color-octet $^3S_{1}^{(8)}$ contribution is strongly suppressed
compared with the others over the entire $P_T$ region considered.
The $^1S_{0}^{(8)}$ -saturated contribution is smaller than the singlet
contribution  where $P_T \leq 5$ GeV.
Integrated over the $P_T$ region ($2.0 \leq P_T \leq 5.0 {\rm GeV}$),
the color-singlet, $ ^1 S_{0}^{(8)}$-saturated,
and $^3 S_{1}^{(8)}$ contributions to $ B \sigma ^{{\rm DPE}}$ are
$0.47$, $0.12$, and $1.4 \times 10^{-2}$ pb respectively.
The total cross section times the leptonic decay branching ratio of
$\jpsi  B$
at $\sqrt{s} = 1.8 {\rm TeV} $, integarated over
the same $ P_T $ and $\eta $ region for DPE, SD, and inclusive production
of associated $\jpsi + \gamma $ are $0.60$, $19$, and
$4.0 \times 10^2 $ pb respectively.
The ratios of the total DPE and SD cross sections  to that of inclusive
production in the central region are
$R^{{\rm DPE}} = 1.5 \times 10^{-3}$ and
$R^{{\rm SD}} = 4.8 \times 10^{-2}$ respectively.

\par
In Fig. 2, we show the $P_T$ distribution of $\jpsi$,
$ B d\sigma / d P_T$, integrated over
the same pseudo-rapidity region at the CERN LHC energy $\sqrt{s} = 14$ TeV
for the inclusive, SD, and DPE processes,
the code for the lines is the same as Fig. 1.
Integrated over the $P_T$ region ($2.0 \leq P_T \leq 5.0 {\rm GeV}$),
the color-singlet, $ ^1 S_{0}^{(8)}$-saturated,
and $^3 S_{1}^{(8)}$ contributions to $ B \sigma ^{{\rm DPE}}$ are
$1.2$, $0.30$, and $3.6 \times 10^{-2}$ pb respectively.
The total cross section times the leptonic decay branching ratio of
$\jpsi  B$  at $\sqrt{s} = 14 {\rm TeV} $, integarated over
the same $ P_T $ and $\eta $ region for DPE, SD, and inclusive production
of associated $\jpsi + \gamma $ are $1.6$, $88$, and
$3.1 \times 10^3 $ pb respectively.
The ratios of the total DPE and SD cross sections  to that of inclusive
production in the central region are
$R^{{\rm DPE}} = 5.2 \times 10^{-4}$ and
$R^{{\rm SD}} = 2.8 \times 10^{-2}$ respectively.
We have varied the color-octet matrix elements
$\moso + \frac{3.5}{m_c^2} \mtpo $ and $\mtso$
by multiplied them by a factor between
$1/10$ and $ 2 $,
the ratios $R^{{\rm DPE}}$ and $R^{{\rm SD }}$  above are unvaried.
This character demonstrates that the ratios $R^{{\rm DPE }}$ and
$R^{{\rm SD }}$ are insensitive to the values of color-octet matrix elements.

\par
In TABLE I., we show the ratios
\ba
R^{{\rm DPE}}(P_T) &=& \frac{ d \sigma ^{{\rm DPE}} }{d P_T} /
\frac{d \sigma ^{{\rm Inclusive}} } {d P_T }, \nonumber \\
R^{{\rm SD}}(P_T) &=& \frac{ d \sigma ^{{\rm SD}} }{d P_T} /
                     \frac{d \sigma ^{{\rm Inclusive}} } {d P_T }  
\ea
at $\sqrt{s} = 1.8 $ and $14$ TeV in the central rapidity region.

%%%%%%%% Insert TABLE I here. %%%%%%%%%%%%%%%%%%%%%%%%%%%%%%%%%%%%%%%%%%%%%%%

\par
From this table, we can see that $R^{{\rm SD }}(P_T)$ is almost constant
for the $P_T$ region considered, so $ R^{{\rm SD }}$ is insensitive to
the $P_T$ smearing effects.
But $R^{{\rm DPE }}(P_T)$ varies rapidly with $P_T $, from $2$ GeV to
$5$ GeV, which decreases by almost one order, so
$R^{{\rm DPE }}$ is sensitive to the $ P_T$ smearing effects, this
can be attributed to the limited production phase space and an additional 
hard gluon distribution of the Pomeron in the DPE compared with the SD case.

\par
In the above calculation, we use the standard Pomeron flux factor (
valid at small $t$ ) that appears in the triple-Pomeron amplitude for SD.
In order to preserve the shapes of the $ M^2 $ and $t$
distribution in soft single diffraction and predict correctly the
experimentally observed SD cross section at all energies in $ p - {\bar p}$
collisions, Goulianos\cite{goulianos} proposed to renormalize the Pomeron
flux in an energy-dependent way:
\beq
f^{{\rm RN}}_{{\rm I\!P}/p }(\xi,t) = D f_{{\rm I\!P}/p }(\xi,t),
\eeq
the renormalization factor $ D $ is defined as
\beq
\label{dfactor}
D = {\rm min} (1, \frac{1}{N})
\eeq
with
\beq
\label{nv}
N = \int_{\xi_{{\rm min} }}^{\xi_{{\rm max}}} d \xi \int^{0}_{- \infty} d t
f_{{\rm I\!P}/p }(\xi,t) ,
\eeq
where $\xi_{{\rm min}} = M_0^2/s $ with $M_0^2 = 1.5 {\rm GeV}^2 $(effective
threshold) and $\xi_{{\rm max}} = 0.1 $(coherence limit).
For $ \sqrt{s} = 1.8, 14$ TeV, the renormalization factor
$D = 0.11, 5.2 \times 10^{-2}$ respectively. Furthermore, by comparing the
CDF data on diffractive W production with predictions using the Pomeron
structure function measured  at HERA, Goulianos concluded that the
breakdown of hard diffractive factorization in hadron-hadron collisions is
due to the breakdown of the Regge factorization already observed in soft
diffraction\cite{gouln2}.
In this picture, the renormalization factor D indicates
the factorization broken effects and takes the role of the
survival probability for hadron emerges from the diffractive collision
intact\cite{soper}.
Using the renormalized Pomeron flux factor,
the $P_T$ distribution
calculated above must be multiplied by a factor $D^2$ and $D$ for DPE and
SD case respectively,
therefore, the ratio $R^{{\rm DPE}}$ and $R^{{\rm SD}}$  are
proportional to $ D^2 f_g^2 $ and $ D f_g $ respectively , hence  they
are sensitive to the gluon fraction of the Pomeron $f_g$ and the
renormalization factor $ D $ which indicates the factorization  broken
effects. From other diffractive production experiments, $f_g$ can
be determined, the renormalization factor $D$ can
be determined precisely from those ratios, and {\it vice versa}.
So measuring these ratios can probe the gluon density in the Pomeron and
shed light  on the nature of hard diffractive factorization breaking.
Furthermore, sence
$R^{{\rm DPE }}$ is sensitive to the $ P_T$ smearing effects,
the experimental study of associated $\jpsi + \gamma $ production
through DPE can give valuble information about the $P_T$ smearing
effects which are needed to solve the large-$z$ discrepancy seen
by comparing NRQCD predictions with the HERA data on inelastic
$\jpsi$ photoproduction\cite{jpsikt}.

\par
Experimentally, the nondiffractive background to the diffractive associated
$\jpsi + \gamma$ production must be dropped out in order to obtain useful
information,
this can be attained by performing the rapid gap analysis or using the
Forward Proton Detector.

\par
In conclusion, in this BRIEF REPORT  we have shown that  
diffractive associated $\jpsi + \gamma $ production at large $P_T$
is sensitive
to the gluon content of the Pomeron and the factorization broken
effects in hard diffraction. Although the diffractive
and inclusive production  cross sections are sensitive to the values of
coler-octet matrix elements, the ratioa of the diffractive to inclusive
$\jpsi + \gamma $ production are not so and
proportional to $D^2 f_g^2 $ and $ D f_g $ for DPE and SD case respectively,
hence they are sensitive to the gluon fraction of the Pomeron  and the
the factorization  broken
effects. So experimental measurement of
these ratios at the Tevatron and LHC can shed light on the nature of Pomeron
and hard diffractive factorization breaking.
Furthermore, sence
the ratio of the total DPE cross section  to that of inclusive
production is sensitive to the $ P_T$ smearing effects,
the experimental study of associated $\jpsi + \gamma $ production
through DPE can give valuble information about the $P_T$ smearing
effects.

\vskip 1cm
\begin{center}
{\bf\large Acknowledgments}
\end{center}
This work is supported in part by the National Natural Science Foundation of
China, Doctoral Program Foundation of Institution of Higher Education of
China and Hebei Natural Province Science Foundation, China.

\newpage
\centerline{\bf \large Figure Captions}
\vskip 2cm
\noindent
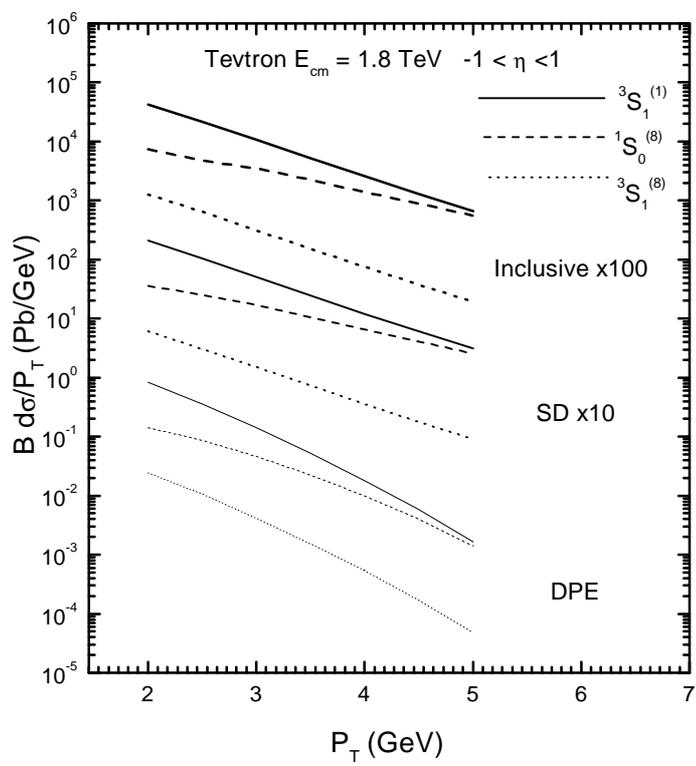
Fig.~1. Transverse momentum of $\jpsi (P_T) $ distribution
$ B d\sigma /dP_T $,
integrated over the $\jpsi$ pseudo-rapidity range $|\eta| \leq 1 $
(central region), for associated $\jpsi + \gamma $ production through
double Pomeron exchange (lower), in single diffractive (middle), and
inclusive (upper) processes at the Tevatron ($\sqrt{s} = 1.8~{\rm TeV}$).
Here $B$ is
the branching ratio of $\jpsi \to \mu^+ \mu^-  (B = 0.0594)$.
The solid line is the color-singlet gluon-gluon fusion contribution,
the dashed line represents $ ^1 S_{0}^{(8)}$-saturated color-octet
contribution, the dotted one is $^3S_{1}^{(8)}$ color-octet contribution.
The inclusive and single production data are multiplied by a factor
100 and 10 respectively.
\noindent
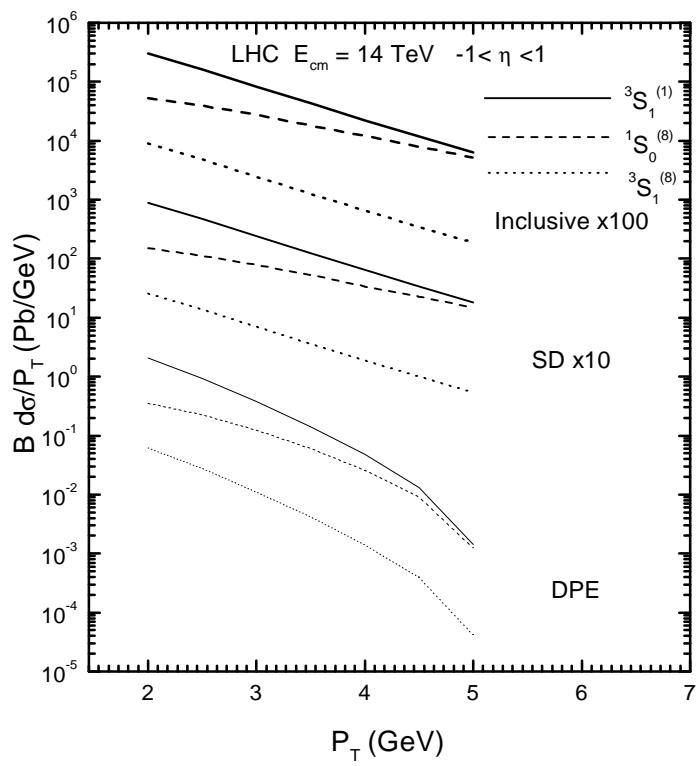
Fig.~2. Transverse momentum of $\jpsi (P_T) $ distribution
$ B d\sigma /dP_T $,
integrated over the $\jpsi$ pseudo-rapidity range $|\eta| \leq 1 $
(central region), for associated $\jpsi + \gamma $ production through
double Pomeron exchange (lower), in single diffractive (middle), and
inclusive (upper) processes at the CERN LHC 
($\sqrt{s} = 14~{\rm TeV}$). Here $B$ is
the branching ratio of $\jpsi \to \mu^+ \mu^-  (B = 0.0594)$.
The solid line is the color-singlet gluon-gluon fusion contribution,
the dashed line represents $ ^1 S_{0}^{(8)}$-saturated color-octet
contribution, the dotted one is $^3S_{1}^{(8)}$ color-octet contribution.
The inclusive and single production data are multiplied by a factor
100 and 10 respectively.

\newpage
\begin{table}[tbp]
\caption{ The ratio $R^{{\rm SD}} (P_T)$, $R^{{\rm DPE}} (P_T)$
as a function of $P_T$ at the Tevatron  and LHC energies 
in central region $(-1 \leq \eta \leq 1)$ using the standard Pomeron flux
(D = 1.0). }
\begin{center}
\begin{tabular}{|cccccccc|}
\hline \hline
$P_T$ (GeV)          &  2.0   &  2.5   &  3.0  &  3.5  &  4.0   & 4.5 & 5.0 \\
\hline
$R^{{\rm SD}} ( P_T )~~ ( \sqrt{s} = 1.8 ~{\rm TeV} ) ~~( 10^{-2} ) $ &
4.9 & 4.9 & 4.8 & 4.8 & 4.7 & 4.7 & 4.7 \\
$R^{{\rm SD}} ( P_T )~~ ( \sqrt{s} = 14 ~~{\rm TeV} ) ~~( 10^{-2} ) $  &
2.9 & 2.9 & 2.9 & 2.9 & 2.9 & 2.9 & 2.9 \\
\hline 
$R^{{\rm DPE}} ( P_T )~ ( \sqrt{s} = 1.8 ~{\rm TeV} ) ~( 10^{-3} ) $ &
2.0 & 1.7 & 1.3 & 1.0 & 0.71 & 0.46 & 0.25 \\
$R^{{\rm DPE}} ( P_T )~ ( \sqrt{s} = 14 ~~{\rm TeV} ) ~( 10^{-4} ) $  &
 6.9 & 5.8 & 4.5 & 3.3 & 2.2 & 1.1 & 0.23 \\
\hline \hline
\end{tabular}
\end{center}
\end{table}


\begin{references}

\bibitem{collins} P. D. B. Collins, {\it An introduction to Regge theory and
High Energy Physics }, Cambridge University Press, Cambridge, England,
1977;
K. Goulianos, Phys. Rep. {\bf 101}, 169 (1985).

\bibitem{ingelmam} G. Ingelman and P. E. Schlein, Phys. Lett. B {\bf 152},
    256 (1985). 

\bibitem{ua8} UA8 Collaboration, A. Brandt {\it et al.}, Phys. Lett. B
{\bf 297}, 417 (1992) ; R. Bonino {\it et al.}, {\it ibid.} {\bf 211},
239 (1988). 

\bibitem{zeus} ZEUS Collaboration, M. Derrick {\it et al.}, Z. Phys. C
{\bf 68}, 569 (1995) ; Phys. Lett. B {\bf 356}, 129 (1995).

\bibitem{h1} H1 Collaboration, T. Ahmed {\it et al.},  Phys. Lett. B
{\bf 348}, 681 (1995) ; C. Adloff {\it at al.}, Z. Phys. C {\bf 76},
613 (1997);
For a recent review on diffraction at HERA, see
P. Marage, hep-ph/9810551.

\bibitem{cdfwj} CDF Collaboration, F. Abe {\it et al.},  Phys. Rev. Lett.
{\bf 78}, 2698 (1997); {\bf 79}, 2636 (1997).

\bibitem{jcollins} J. C. Collins, Phys. Rev. D {\bf 57}, 3051 (1998).

\bibitem{copom} J. C. Collins, L. Frankfurt, and M. Strikmam,
Phys. Lett. B {\bf 307}, 161 (1993);
A. Berera and D. E. Soper, Phys. Rev. D {\bf 50}, 4328 (1994).

\bibitem{testa} L. Alvero, J. C. Collins, J. Terron, and
J. J. Whitmore, hep-ph/9805268;
L. Alvero, J. C. Collins, and J. J. Whitmore, hep-ph/9806340.


\bibitem{xu} Jia-Sheng Xu and Hong-An Peng, Phys. Rev. D {\bf 59} (1), (1999).

\bibitem{nrqcd} W. E. Caswell and G. P. Lepage, Phys. Lett. B {\bf 167},
    437 (1986).

\bibitem{power} G. P. Lepage, L. Magnea, C. Nakhleh, U. Magnea, and
K. Hornbostel, Phys. Rev. D {\bf 46}, 4052 (1992).

\bibitem{cdfjpsi} CDF Collaboration, F. Abe {\it et al.}, Phys. Rev. Lett.
{\bf 79}, 572 (1997); {\bf 79}, 578 (1997).

\bibitem{brt-cho}
For some recent reviews, see
E. Braaten, S. Fleming, and T.C. Yuan, Annu. Rev. Nucl. Part. Sci. {\bf 46},
197 (1996); hep-ph/9602374 ; E. Braaten, hep-ph/9702225; hep-ph/9810390;
M. Beneke, hep-ph/9703429; and references therein. 

\bibitem{nrqcdfs} G. T. Bodwin, E. Braaten, and G. P. Lepage,
Phys. Rev. D {\bf 51}, 1125 (1995); {\bf 55}, 5853(E) (1997).

\bibitem{berg} E. L. Berger and D. Jones, Phys. Rev. D {\bf 23}, 1512 (1981);
            R. Baier and R. Rukel, Z. Phys. C {\bf 19} 251 (1983). 

\bibitem{kim} C. S. Kim, Jungil Lee, and H.S. Song, Phys. Rev. D {\bf 55},
5429 (1997); P. Ko, Jungil Lee, and H.S. Song, {\it ibid.}, {\bf 54}, 4312
(1996).

\bibitem{goulianos} K. Goulianos, Phys. Lett. B {\bf 358}, 379 (1995);
 {\bf 363}, 268(E) (1995) ;
 K. Goulianos and J. Montanha, hep-ph/9805496.

\bibitem{beneke} M. Beneke and M. Kr$\ddot{a}$mer, Phys. Rev. D {\bf 55},
R5269 (1997).

\bibitem{grv} M. Gl$\ddot{u}$ck, E. Reya, and A. Vogt, Z. Phys. C {\bf 67},
433 (1995). 

\bibitem{gouln2} K . Goulianos, hep-ph/9708217.

\bibitem{soper} D. E. Soper, hep-ph/9707384.

\bibitem{jpsikt} H1 Collaboration, S. Aid {\it et al.}, Phys. Lett. B
{\bf 472}, 3 (1996); ZEUS Collaboration, J. Breitweg {\it et al.},
hep-ex/9708010;
B. Cano-Coloma and M. A. Sanchis-Lozano, Nucl. Phys. {\bf B508}, 753
 (1997); B. A. Kniehl and G. Kramer, hep-ph/9803256;
K. Sridhar, A. D. Martin, and  W. J. Stirling, hep-ph/9806253.

\end{references}
\end{document}